\documentclass[structabstract]{aa} 
\usepackage{graphicx}
\usepackage{txfonts}
\usepackage{natbib}
\newcommand\be{\begin{equation}}
\newcommand\ee{\end{equation}}
\newcommand\bea{\begin{eqnarray}}
\newcommand\eea{\end{eqnarray}}

\usepackage{color}

\newcommand{\refe}{}

\graphicspath{{./Figures/}}

\begin{document}
   \title{Flux Modulation from the Rossby Wave Instability in microquasars' accretion disks: toward a HFQPO model}
   \titlerunning{Rossby wave instability: toward a HFQPO model}

   \author{F. H. Vincent
          \inst{1}   
          \and
          H. Meheut
          \inst{2,3}       
          \and
          P. Varniere\inst{1}
          \and 
          T. Paumard\inst{4}
          }

   \institute{AstroParticule et Cosmologie (APC), Universit\'e Paris Diderot, 10 rue A. Domon et L. Duquet, 75205 Paris Cedex 13, France \\
             \email{frederic.vincent@obspm.fr}
         \and
         Physikalisches Institut, Universit\"at Bern, 3012 Bern, Switzerland
         \and 
         CEA, Irfu, SAp, Centre de Saclay, F-91191 Gif-sur-Yvette, France\\
         \email{heloise.meheut@cea.fr}
         \and
        	LESIA, Observatoire de Paris, CNRS, Universit\'e Pierre et Marie Curie, Universit\'e Paris Diderot, 5 place Jules Janssen, 92190 Meudon, France\\		
             }

   \date{\today}
 
  \abstract
   {There have been a long string of efforts to understand the source of the variability observed in microquasars, especially concerning the elusive 
   High-Frequency Quasi-Periodic Oscillation.
   These oscillations are among the fastest phenomena that affect matter in the vicinity of stellar black holes 
   and therefore could be used as probes of strong-field general relativity. Nevertheless, no model has yet gained wide acceptance.}
   {The aim of this article is to investigate the model derived from the occurrence of the Rossby wave instability at the inner edge of the accretion disk. 
   In particular, our goal here is to demonstrate the capacity of this instability to modulate the observed flux \refe{in agreement with the}
   observed results.}
   {We use the AMRVAC hydrodynamical code to model the instability in a 3D optically thin disk. The GYOTO ray-tracing code is then used to 
   compute the associated light curve.}
   {We show that the 3D Rossby wave instability is able to modulate the flux \refe{well within the observed limits.}
   We highlight that 2D simulations allow \refe{us} to obtain the same general characteristics of the light curve as 3D calculations. With the time resolution we adopted in this work, three dimensional simulations do not give rise to any new observable features that could be \refe{detected} by current instrumentation or archive data.}
   {}


   \maketitle


\section{Introduction}
\label{sec:intro}

\refe{Black hole binary systems exhibit a wide range of variability patterns. Among these, the high frequency quasi-periodic oscillations (HFQPOs) }appear as \refe{small} narrow peaks in the power density spectrum. 
Their frequencies, of several tens to a few hundred Hertz, \refe{correspond} 
to the rotation frequency in the inner part of the disk surrounding the black hole (see more details in the review of \citealt{VDK06}) \refe{making them an interesting tool to study strong-field gravity.}
Even though they are the fastest phenomena observed in the vicinity of black holes, HFQPOs are also extremely elusive, especially when compared to the lower frequency quasi-periodic oscillations (LFQPO).  
For this reason, we do not have a rich domain of observations from which to draw constraints for the models. 
This has led to a wide variety of models  (see e.g. \citealt{LAI09} for a very brief review) focusing on one or another \refe{of the} observed characteristics of the HFQPOs. 
As  more and more data \refe{are obtained}, especially with the next generation telescopes such as LOFT~\citep{BOZ11}, we will be able to {\refe{gain} a better picture of what a model must explain. 
Nevertheless, we already can make a stringent list of constraints based on the eight objects with recurrent HFQPOs~\citep[see for example the review by][]{REM06}. 
Among the most important of those constraints we note:
\begin{description} 
\item[1-] The emitted flux is modulated at a level of a few percent in X-ray and this modulation increases with energy. 
                 This first and fundamental point, often not discussed by models, is the aim of this paper.
                 
\item[2-] The frequency of the modulation has a small but significant range that must be explained.

\item[3-] The HFQPOs occur sometimes, but not always, in pairs close to a $2$:$3$ ratio. Those occurrences, and not just the close ratio, need to be elucidated. 

\item[4-] Another point often neglected in models for HFQPOs is the fact that, sometimes, they co-exist with the more ubiquitous LFQPOs.
\end{description}

      The model we are investigating here is based on the existence of the Rossby wave instability (RWI) 
      at the inner edge of the disk as was discussed in \citet{TAV06}.
       Indeed, it was shown that the existence of an innermost stable circular orbit (ISCO) around a black hole makes the disk prone to the RWI
       \refe{because the vorticity profile has a natural extremum}.
       This dynamic instability results in the formation of large-scale spiral density waves and Rossby waves that can reach high amplitudes. 
       Depending on the disk's physical parameters, different modes of this instability will be selected, most often with azimuthal mode number 
       $m=2$ or $3$ \citep{TAV06} which gives a natural explanation for some of the observed characteristics of HFQPOs. 
       \refe{Nevertheless, }the important step of computing the actual flux modulation one would observe when such an instability occurs in a disk \refe{was still missing}.
       Here we focus on this point and therefore restrict ourselves to \refe{ the case of the} HFQPO alone. 
      Indeed, we are trying to explore the capacity of one
      particular HFQPO model to modulate the X-ray flux and having only one instability in the disk makes the results more conspicuous. 
      To this end we will use the purely hydrodynamic version of the Rossby wave instability (RWI, \citealt{LOV99}), without the extra destabilizing 
      effect provided by a poloidal magnetic field  \citep{TAV06} or the extra stabilizing effect of a toroidal magnetic field \citep{YL09}. 
      
       This is the continuation of the long string of studies on the impact of the RWI in disks such as was done in the context of protoplanetary disks 
       \citep{RJS12} and the galactic centre \citep{FAL07}. 
       Whereas the first case is highly different from ours, the second is more comparable and we use similar tools, but \citet{FAL07} restricted 
       themselves to the 2D case and to the special context of the galactic centre source Sgr~A*.

     Our main interest here will thus be to answer the key question: is the RWI able to modulate the flux of a microquasar at a level coherent with observations?      
     In a more practical manner, we will also address the question of the necessity  of  full 3D simulations of the RWI, 
     and to what extent it is sufficient to restrict oneself to 2D simulations. Those 2D simulations are more accessible in terms of computing and 
     storage resources,  therefore allowing us to do much larger parameter studies and with a higher time resolution that could be compared 
     with future LOFT observations. 
     These questions will be addressed by computing the light curve of a microquasar subject to the RWI, as observed by a distant observer. 
     This computation will be performed by means of a general relativistic ray-tracing code~\citep{VIN11}. 
     The main general relativistic effect that affects the RWI being the existence of an ISCO, we use a pseudo-Newtonian potential in order to 
     model this in the hydrodynamical simulations, instead of considering the full general relativistic case. 
    This simplifying assumption is sufficient in order to derive a proof of principle of the ability of the RWI to modulate the flux of microquasars.

Sect.~\ref{sec:simusamr} briefly presents the RWI and then develops the hydrodynamical simulations that allow us to model the 
unstable accretion structure. The following section is focused on the ray-tracing procedure used to compute observable quantities. 
Sect.~\ref{sec:results} presents the light curves that were obtained and Sect.~\ref{sec:conclu} gives conclusions and prospects.


\section{Hydrodynamical simulation of the RWI }
\label{sec:simusamr}

The RWI has been discussed in multiple astrophysical contexts, from the galactic centre to  protoplanetary disks. 
It can be seen as the form the Kelvin-Helmholtz instability takes in differentially rotating disks, and has a similar instability criterion.
For two-dimensional (vertically integrated) barotropic disks, the RWI can be triggered provided an extremum exists in a quantity $\mathcal L$ which 
is {the inverse vortensity\footnote{Vortensity is the ratio of vertical vorticity to surface density.}}:

\begin{equation}
\mathcal L = \frac{\Sigma \Omega}{2 \kappa^{2}}\frac{p}{\Sigma^{\gamma}},
\label{eq:vortensity}
\end{equation}
where $p$ is the pressure, $\Sigma$ is the surface density, $\Omega$ is the rotation frequency, $\kappa^{2}=2\Omega/ r \, \mathrm{d}/\mathrm{d}r \left(r^2\Omega\right)$ is the squared epicyclic frequency and $\gamma$ is the adiabatic index.
An extremum of vortensity can thus typically arise from an extremum of the epicyclic frequency\refe{, as is the case close to the ISCO,} or from an extremum in the density profile 
\refe{which} gives rise to the RWI in the disk. Once it is triggered, it leads to large scale spiral density waves and Rossby vortices.  
The dominant mode (i.e. the number of vortices) is dependent on  the disk conditions  \citep{TAV06}.

\subsection{The RWI as a model for the HFQPO}

Due to general relativistic effects, the epicyclic frequency profile will show a maximum  in the vicinity of the black hole's ISCO, creating a vortensity extremum .  
\citet{TAV06} showed that it would lead to the growth of the RWI at the inner edge of black hole binaries when the disk \refe{nears} its ISCO. 
From these facts, it seems natural that the RWI model may be capable of accounting for points 2- to 4- in the Introduction above (that will not be investigated in detail here). 

\begin{description} 
\item[2-] 
	  The frequency range:
           the exact location of the resulting vortensity extremum will depend on the density profile of the disk on top of the epicyclic frequency.
           Thus the location of the launching of the instability can vary depending on the disk properties. As the Rossby vortices will orbit around
           the central black hole with the disk's rotation frequency at the radius of launching, this variation of the disk properties gives rise to a range 
           of different possible frequencies of the signal.
            
\item[3-] 
               Pairs of frequencies: different modes can dominate the signal depending on the physical state of the disk. This was adressed in \citet{TAV06} 
               and more recently in~\citet{VAR12b}. The 2:3 frequency pairs can be naturally explained by a superposition of azimuthal modes $m=2$ and $m=3$.

\item[4-] 
              Coexistence of HF- and LFQPO: we have tackled this point by showing the ability of the RWI to co-exist with another instability that we proposed to be at the origin 
             of the LFQPO~\citep{VAR12a}, the Accretion Ejection Instability (AEI). A disk giving rise to both AEI and RWI will thus \refe{naturally} show both types of QPOs.
\end{description}

     It seems therefore natural to address in more detail if the RWI is strong enough to give rise to a modulation that could explain the observations.

\subsection{Numerical setup}
 \label{sec:param}

To perform the hydrodynamical simulation of the RWI we use the 
Message Passing Interface-Adaptive Mesh Refinement Versatile Advection Code (MPI-AMRVAC) developed by~\citet{KEP11}. 
The numerical scheme is the same for all refinement levels, namely the Total Variation Diminishing Lax-Friedrich scheme \citep[see][]{TOO96} 
with a third order accurate Koren limiter \citep{KOR93} on the primitive variables. 

We consider a geometrically thin disk  and neglect self-gravity. The Euler equations in cylindrical coordinates $(r,\varphi, z)$ read:
\begin{eqnarray}
\partial_t\rho+\vec \nabla\cdot (\vec v\rho)=0 \,,\\
\partial_t(\rho\vec v)+\vec\nabla\cdot(\vec v\rho\vec v)+\vec\nabla p=-\rho\vec\nabla \Phi_G \,,
\end{eqnarray}
where $\rho$ is the mass density of the fluid, $\vec v$ its velocity, and $p$ its pressure. 
We consider a barotropic flow, i.e. the entropy is constant in the entire system. 
The pressure is then $p = S\rho^\gamma$, with the adiabatic index $\gamma= 5/3$ and the constant $S$ related to entropy. 
The sound speed is given by $c_s^2=\gamma p/\rho=S\gamma \rho^{\gamma-1}$ and the temperature by $T\propto p/\rho \propto S\rho^{\gamma-1}$. 
All the distances are normalized by the ISCO radius $r_0$. The same applies for temperatures normalized at ISCO by: $T_{0}=10^7K$, while all times are 
in units of the ISCO period $t_{\mathrm{ISCO}}$. 
\newline

The general relativistic effects have to be taken into account to study the innermost region of the black hole disk. 
In order to provide a proof of principle of the ability of the 3D RWI to modulate the flux of microquasars, we use the minimum model \refe{necessary} to mimic some of these general relativistic effects by using a pseudo-Newtonian potential $\Phi_G$. For our simulations, we use the common \citet{PAC80} gravitational potential:
\be
\Phi_G=-\frac{GM_*}{\sqrt{r^2+z^2}-2\,r_g}
\label{eq:PW}
\ee
where $r_g=\frac{GM_*}{c^2}$ is the gravitational radius, $G$ the gravitational constant, $M_*$ the black hole mass, $c$ the speed of light in vacuum. 
With this gravitational potential, one can define the inner limit below which there is no stable circular orbit, the ISCO, located at $r_0=6\,r_g$. Let us stress the fact that this choice of potential allows us to mimic only some aspects of the Schwarzschild metric. The effect of the black hole spin is not taken into account. It is also important to keep in mind that this pseudo-Newtonian potential, although widely used, is far from giving an exact description of general relativity. However, the only characteristic of the Schwarzschild metric that is of primordial importance as far as RWI is concerned is the existence of an ISCO, and this is correctly modeled by Eq.~\ref{eq:PW}.

Let us mention that we have also developed 2D RWI simulations with alternative pseudo-Newtonian potentials that take into account the rotation of the black hole. These alternative potentials were taken from~\citet{artemova96}. The RWI also develops  well in these potentials, and future works will be dedicated to studying the effect of the black hole's rotation on the observables.

\subsection{Disk setup and boundary  conditions}

The grid is cylindrical \refe{with the unit length}  defined as the ISCO radius $r_0$. The radial coordinate $r$ is in the range $[0.8,6]$, 
the full azimuthal coordinate spans $[0,2\pi]$ and the vertical coordinate $z$ lies in the range $[0,0.2]$ for the 3D simulations. 
For the fluid simulations, we considered only the upper part of the disk  as the mid-plane is a symmetry plane for the RWI \citep{MEH10,MYL12}. 
For the ray-tracing computations, the full disk is considered by symmetrizing the upper part. 
We used a fixed and homogeneous grid with a resolution $n_{r}\times n_{\varphi} = (384,72)$ in 2D and 
$n_{r}\times n_{\varphi} \times n_z= (384,72,72)$ in 3D, which is slightly higher than the resolution used in~\cite{MEH10}.

\refe{As we describe gravitation by the pseudo-Newtonian potential in Eq.~\ref{eq:PW},
the epicyclic frequency will show a maximum at a radius slightly higher than the  ISCO radius $r_0$. 
This epicyclic frequency maximum gives rise to an extremum of the initial vortensity profile, which is shown in Fig.~\ref{fig:criterium}. The existence of this vortensity extremum does not depend on the choice of initial density profile even though it does influence its precise location
in the disk.} 
\refe{Considering the density, we choose a typical power-law profile 
with a slope of $-3/2$. This choice, while reasonable, allows us to obtain a steeper vortensity extremum  and hence a stronger instability. 
Note that the precise value of the rms of the modulation depends on this choice of slope. 
However, we are here only interested in proving the existence of the flux modulation, not comparing a precise value of modulation to observations. 
If future instruments give better data on the rise of the HFQPO we might be able to use these to constrain the disk model that could reproduce the observed data. }

\refe{The initial density profile then reads:}



\be
\rho_M\equiv\rho(r,\varphi,z=0)=\frac{\rho_0}{2}\Big[1+\tanh\big(\frac{r-r_B}{\sigma }\big)\Big]\left(\frac{r}{r_{0}}\right)^\alpha,
\label{eq:rhomidplane}
\ee

the hydrostatic equilibrium gives the vertical profile of the density
\be
\rho(r,\varphi,z)=\rho_M\Big(1-\frac{\gamma-1}{\gamma S\rho_M^{\gamma-1}}\bigg[\frac{GM_*}{r-2\,r_g}-\frac{GM_*}{\sqrt{r^2+z^2}-2\,r_g}\Big]\bigg)^{\frac{1}{\gamma-1}}
\label{eq:rhoini}
\ee
where $r_B=1.3\,r_0$ gives the position of the density maximum \citep[its value was choosen to fit the simulations of][]{TAV06},
$r_g=\,r_0/6$ is the gravitational radius, $\sigma=0.05\,r_0$ gives the width of the 
plunging region, $\alpha=-3/2$ is the density slope, and $\rho_0$ is the minimum density of the simulation. 
\refe{These parameters, and mainly the choice of density slope, can modify the structure and characteristics of the RWI. 
{\bf In our case, the existence of an extremum in $\kappa$ limits the impact of the density power-law index and any reasonable choice 
will exhibit the instability.}
The detailed effect of the initial density profile on the growth rate of the instability and its saturation level will be studied in a forthcoming paper (Meheut, Lovelace, Lai, 2013) but will not influence its ability to modulate the flux.}


We also considered a 2D disk with the surface density  defined as $\Sigma\propto r^{-3/2}$ in the outer region. The initial conditions of the 2D and 3D simulations are then highly different.
In both cases, the azimuthal velocity is determined by force balance:

\be
v_\varphi=\sqrt{\frac{GM_*r}{(r-2\,r_g)^2}+\alpha S\gamma \rho_M^{\gamma-1}+S\gamma \rho_M^{\gamma-2}\frac{(r/r_0)^{\alpha+1}\rho_0}{2\sigma\cosh^2{\frac{r-r_B}{\sigma }}}}.
\label{eq:vphini}
\ee


\begin{figure}
	\centering
	\includegraphics[height=6cm,width=7.5cm]{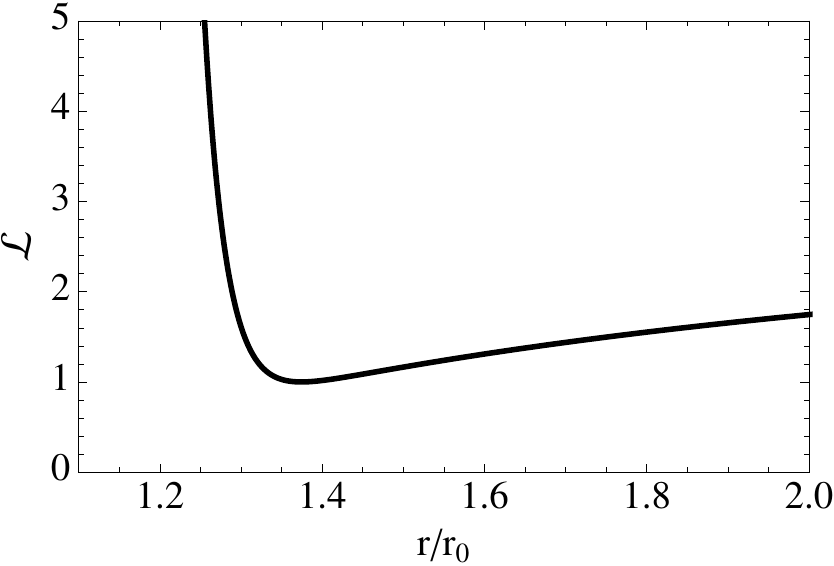}
	\caption{Inverse vortensity profile, normalized to its extremum value. The instability is triggered at the location of the extremum, i.e. at $r \approx 1.4 \,r_0$.}
	\label{fig:criterium}
\end{figure}

The density and velocity profiles given in Eqs.~\ref{eq:rhoini}-\ref{eq:vphini} include high amplitude gradients and do not give exact numerical equilibrium. For this reason, the initial conditions of our simulations are numerically computed from this pseudo-equilibrium. This means that a first simulation is done without any perturbations and is run until the disk  has reached a permanent stage which is chosen as the initial conditions. This disk is then perturbed with small amplitude ($\sim 10^{-3}\,v_{\varphi}$) random perturbations on the radial velocity which act as seeds for the instability.

The inner boundary condition is a no-inflow condition and the outer boundary condition is a null radial velocity condition. 
\citet{MYL12} have shown that the boundary conditions do not change significantly the growth rate of the RWI due to its confinement in the vortensity bump. 
Moreover the inner edge of the simulation being inside the ISCO, any reflected wave would not reach the region of interest for the instability. 
Therefore, the boundary conditions are not determinant for these simulations. 
For the 3D simulations, symmetric boundary conditions are implemented at the mid-plane, and we use a null vertical velocity at the grid upper boundary limit, situated outside of the disk.

\subsection{Time evolution of the RWI } 

Fig.~\ref{fig:midplane} shows the density map of the $z=0$ plane of a 3D RWI simulation, when the instability is completely developed. This section presents the time evolution of the instability (at 2D and 3D) from its launch to its complete development.
       
 \begin{figure}
	\centering
	\includegraphics[height=8cm,angle=90,trim=2cm 3.5cm 2.cm 3.4cm,clip=true]{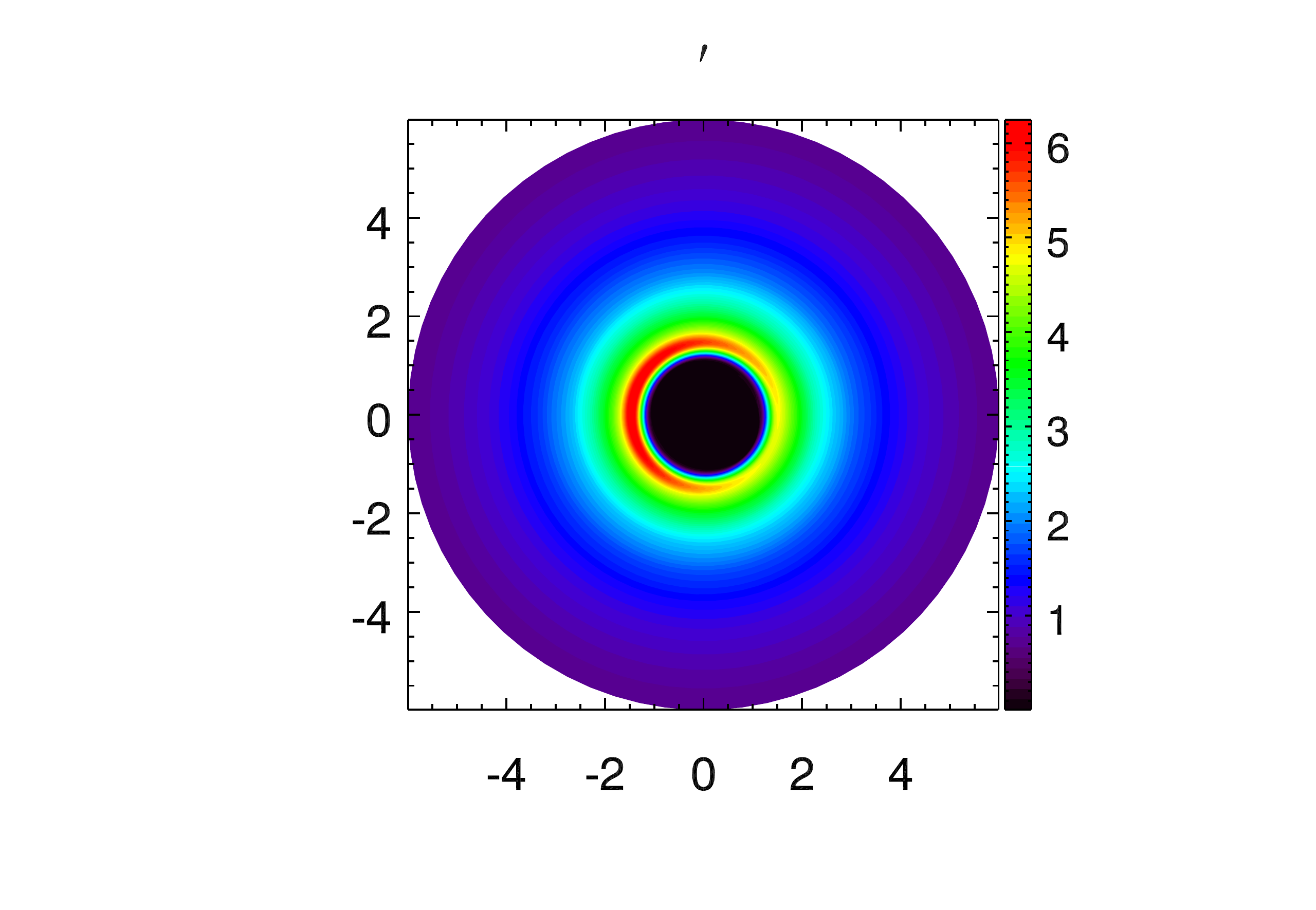}
	\caption{Midplane density map of the 3D simulation when the m=1 mode dominates. The axes are in units of the ISCO radius.}
	\label{fig:midplane}
\end{figure}

At first, the disk is assumed to be 2D and the vertical structure of the disk is neglected. 
The growth of the instability is shown on Fig.~\ref{fig:growth} where the time evolution of the perturbation of density is plotted on a logarithmic scale. 
This allows \refe{the identification of} the linear phase of the instability when the amplitude of the perturbations grows exponentially (thus linearly in logarithm), \refe{as well as} the saturation which is due to non-linearities. 
In the 2D case, the initial conditions 
give rise to a rapid growth of the instability on a time scale of a few periods at ISCO as can be seen in Fig.~\ref{fig:growth}.
During the linear phase, the perturbations can be separated in the different azimuthal modes:
\be
X=\sum_{m\in\mathbb{N}}X_m \,\mathrm{sin}(m\omega t-m\phi)\exp{\gamma_m t}
\ee
where $X=\Sigma$ for 2D simulations and $X=\rho$ for 3D. The quantities $\gamma_m$ are the growth rates of each mode, $\omega$ the characteristic frequency of the instability depending on the position of the extremum of $\mathcal L$, $m$ the azimuthal mode number, and $X_m$ the amplitude of mode $m$. 
Therefore, $X_{m=0}$ corresponds to the axisymmetric part of the density and the frequency of the mode $m$ is the multiple $m\omega$ of the frequency of the fundamental mode. 
Fig.~\ref{fig:growth} also shows the time evolution of the amplitude of the strongest modes. 
During the linear phase, the dominant mode is $m=3$, and later on the disk is dominated by the fundamental mode $m=1$ with important contributions from the modes $m=2$ and $3$.
This evolution of the oscillation modes depends on the disk's astrophysical properties: different modes dominate for different initial disk conditions.
\newline

\begin{figure}
	\centering
	\includegraphics[width=10cm,height=7.5cm,angle=0]{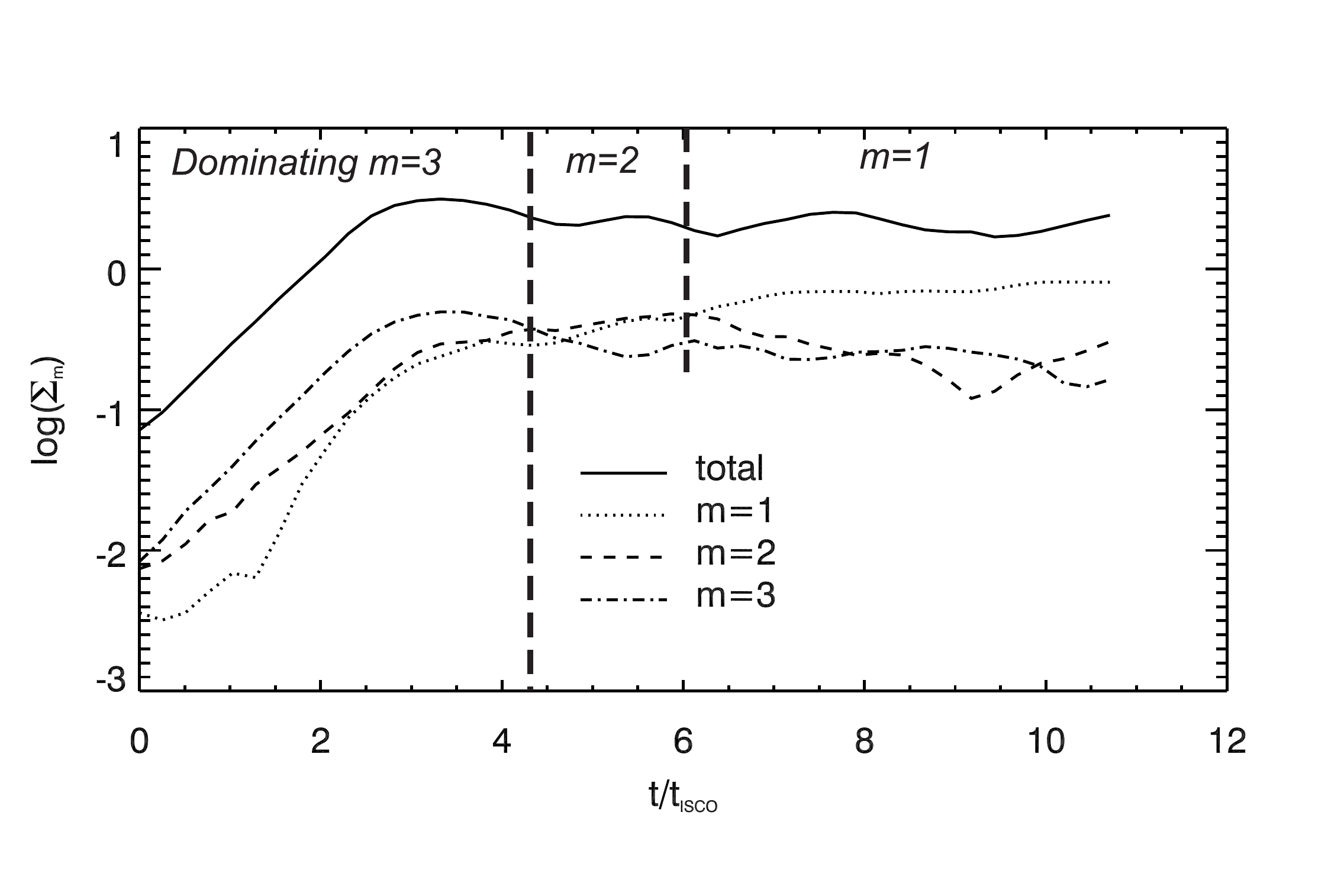}
	\caption{Amplitude of the surface density perturbation in logarithmic scale (solid line) for the 2D simulation.
	The amplitudes of the strongest modes are also plotted in dotted and dashed lines. The vertical dashed lines show the changes of dominant modes, indicated in italic at the top of the figure. The fundamental mode $m=1$ eventually dominates.}
	\label{fig:growth}
\end{figure}

This fact is illustrated in the 3D simulations. Indeed, the initial conditions differ largely between the 2D and 3D simulations, with a different surface density radial profile and absolute value. 
\begin{figure}
	\centering
	\includegraphics[width=9.5cm,height=6cm]{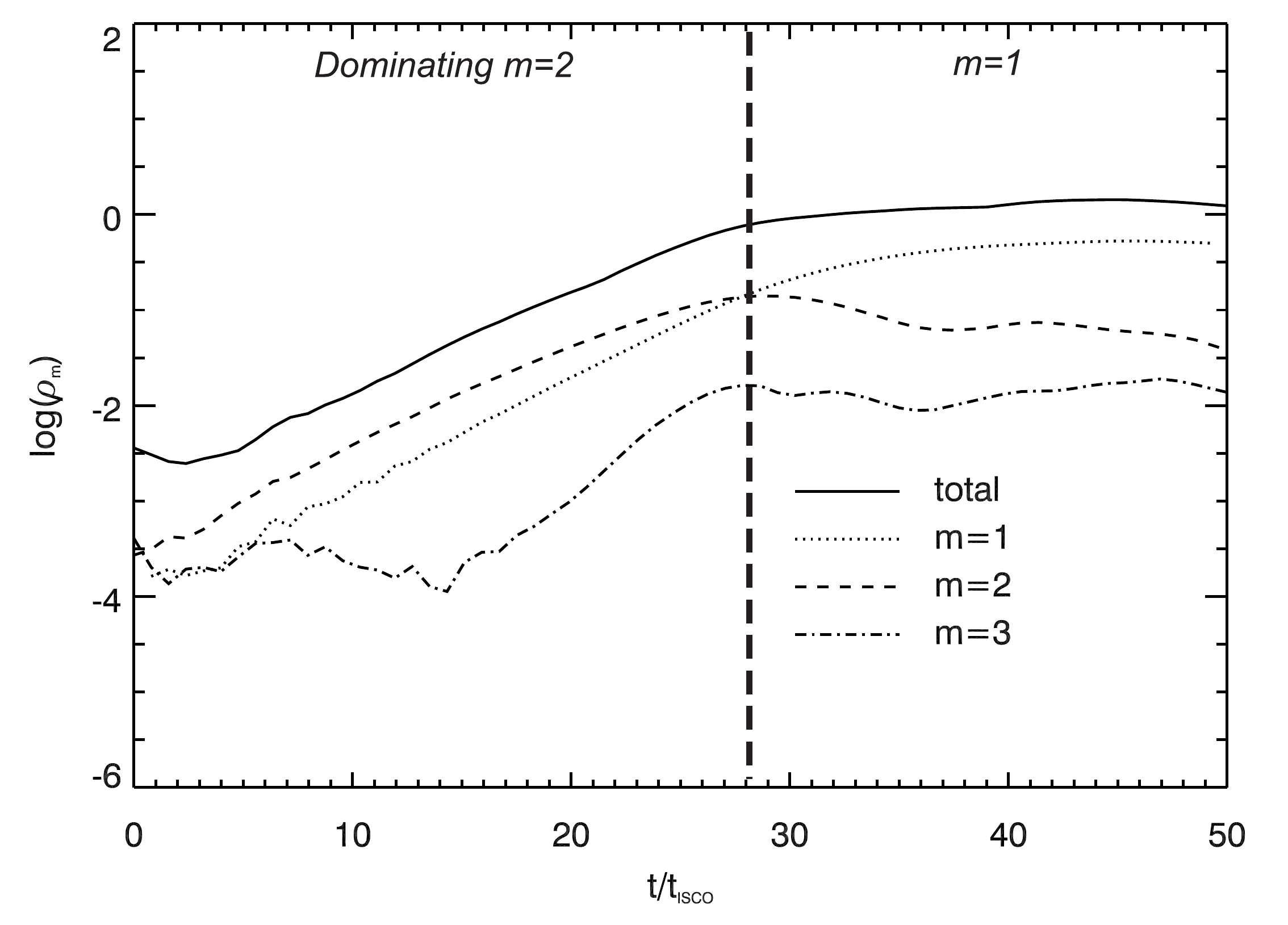}
	\caption{Same as Fig.~\ref{fig:growth} for 3D simulations. However, mind the different time scale.}
	\label{fig:growth3D}
\end{figure}
The characteristic velocity of the waves is then different. This modifies the timescale for the growth of the instability as can be seen in Fig.~\ref{fig:growth3D}. 
This difference is not due to the dimensionality~\citep{MYL12,L12}.  
Nevertheless, the RWI is triggered and saturation of the 3D instability is reached after a few tens of periods at ISCO. 
The dominant modes are also different during the linear phase, but non-linearities still tend to favour the lowest azimuthal mode. 

\refe{
Let us note that, as the epicyclic frequency profile is not evolving with time, the location of the extremum of vortensity will stay the same.
Moreover the vortices grow in a density extremum where they cannot migrate due to the two density slopes of opposite sign. As a consequence, the
Rossby vortices will not migrate during the simulation~\citep[for more details see][]{MKC12}.}

      From the hydrodynamical simulations we obtain the density and velocity at every point of our grid with a sufficient time sampling 
      (of 40 frames in 2D and 6 frames in 3D per orbit at the ISCO) to follow  the RWI in the disk from its rise to its saturation. 
      In the next step those will be used as input for the ray-tracing code in order to trace the impact of the RWI on the observed emission.


\section{Ray-tracing of the RWI with GYOTO}
\label{sec:raytracing}

\subsection{Ray-tracing the accretion disk}

 In order to obtain images, and especially lightcurves, that could be compared with observations we used the general relativistic ray-tracing code GYOTO~\citep{VIN11}.
Null geodesics are computed backward in time in the Schwarzschild metric from a distant observer to the emitting disk. 
The coordinates used by GYOTO are spherical-like and denoted $(\bar{r},\theta,\varphi)$, where $\bar{r}$ is used to differentiate the GYOTO spherical radius from the fluid simulations cylindrical radius $r$.
Once a backward integrated geodesic hits the disk, the density and 3-velocity at the point of emission is linearly interpolated at the time of emission from the results of the MPI-AMRVAC computations. 

\subsubsection{Radiative processes at 2D: blackbody}

The 2D disk is assumed to be optically thick, so the integration is stopped at the first encounter of the disk. The emission is supposed to follow the Planck law: 
the only parameter needed is thus the temperature which can be easily derived from the density according to the following computations.

The gas being assumed ideal:
\be
p = \frac{\rho}{\mu m_u}\,k\,T
\ee
where $\mu$ is the mean molecular weight, $m_u$ is the atomic mass constant and $k$ is the Boltzmann constant. Assuming that the plasma is made of pure hydrogen, $\mu=1$. Using the relation $R=N_A\,k$ between the ideal gas contant $R$, the Avogadro number $N_A$ and the Boltzmann constant together with the expression of the sound speed this yields:
\be
T = \frac{N_A\,m_u}{\gamma}\,\frac{c_{s}^{2}}{R}.
\ee

This allows \refe{the computation of} the emitted specific intensity at the surface of a 2D disk: 
\be
\label{eq:2Demission}
I_{\nu} = B_{\nu}(T)
\ee
where $B_{\nu}$ is the Planck function. 

\subsubsection{Radiative processes at 3D: Bremsstrahlung}

For the three dimensional computations, the only radiative process considered is Bremsstrahlung. As the disk is purely hydrodynamic, there is no synchrotron emission, and Compton scattering is neglected
as we are only interested here in a proof of principle, not in a detailed study of emission processes. The Bremsstrahlung emission is assumed to be thermal, so that the emission coefficient $j_{\nu}$  and absorption coefficient $\alpha_{\nu}$ are related via Kirchhoff's law:

\be
\label{kirch}
j_{\nu}=\alpha_{\nu}\,B_{\nu}.
\ee

The emission coefficient for thermal Bremsstrahlung is given by~\citep{RL79}:

\be
j_\nu = \frac{1}{4\pi}\frac{2^{5}\pi e^{6}}{3m_{\mathrm{e}}c^{3}} \left( \frac{2\pi}{3 k m_{\mathrm{e}}} \right)^{1/2} T^{-1/2} \left(\frac{\rho}{m_{\mathrm{u}}}\right)^{2} \mathrm{exp}\left(-\frac{h\nu}{k T}\right)
\ee
where $e$ is the electron charge, $m_{\mathrm{e}}$ is the electron mass and $h$ is the Planck constant. Here, we assume that the disk is made of pure hydrogen, and that the emission is isotropic in 
the emitter's frame (hence the $1/4\pi$ initial factor). Moreover, the Gaunt factor is neglected as most of the radiation arises from locations in the disk where $h\nu \approx k T$.

Once the emission coefficient is computed, the absorption coefficient is derived by using Eq.~\ref{kirch}.

\subsubsection{Dependency on temperature at 2D and 3D}
\label{compare2D3D}

      Let us investigate the dependency on temperature of the 3D emission process (Bremsstrahlung) as compared to
       the 2D case (blackbody). For temperatures around $10^7$K, from where most of the flux arises,
      the 3D emission coefficient follows $j_\nu^{\mathrm{Br}} \propto T^{2.5} \mathrm{exp}\left(-h \nu / k T\right)$ 
      whereas the 2D specific intensity follows $I_{\nu}^{BB} \propto  \mathrm{exp}\left(-h \nu / k T\right)$. The dependency on 
      temperature is thus much more important in 3D, and the emission arises only from the hottest parts of the disk, whereas the emission
      is more spread out in the 2D case. 

\vspace{0.5cm}

All these computations allow us to determine the specific intensity in the emitter's frame, $I_{\nu_{\mathrm{em}}}$. In order to compute the specific intensity in the observer's frame $I_{\nu_{\mathrm{obs}}}$, the frame invariant quantity $I_{\nu}/\nu^{3}$ is used: $I_{\nu_{\mathrm{obs}}} = \nu_{\mathrm{obs}}^{3}/\nu_{\mathrm{em}}^{3}\,I_{\nu{_\mathrm{em}}}$. The quantity $\nu_{\mathrm{em}}$ can be related to the emitter's 4-velocity. The 3-velocity computed by the fluid simulations is used to determine the emitter's 4-velocity, and the observed specific intensity is thus at hand.

\subsection{Computing the disk image and the light curve}

\begin{figure*}[tbp]
	\centering
	\includegraphics[height=6cm]{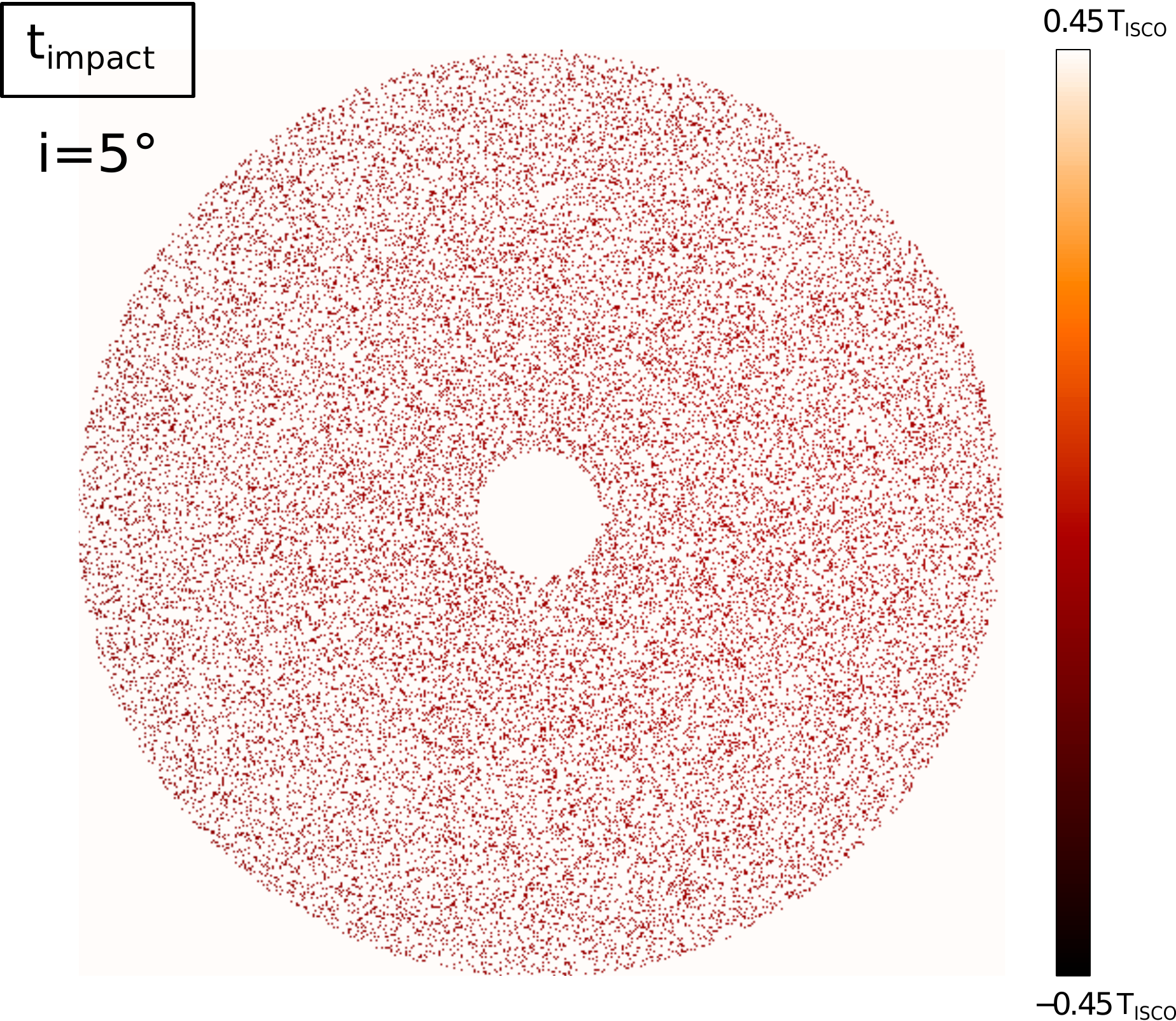}
	\hspace{1cm}
	\includegraphics[height=6cm]{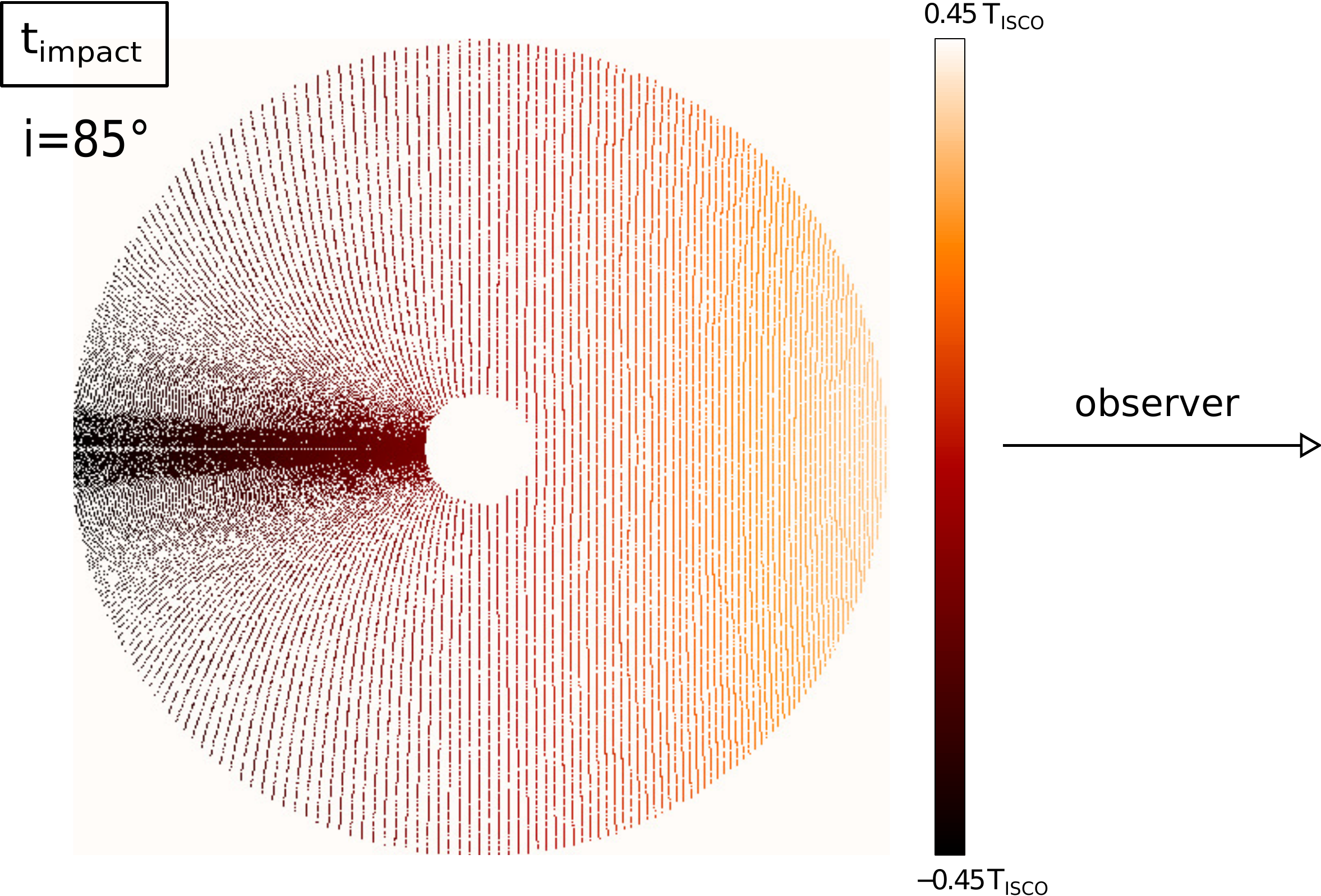}
	\caption{Emission dates of photons at the footpoint of null geodesics connected to the observer's screen, plotted with the same color scale for an inclination of 5$^{\circ}$ (left) and 85$^{\circ}$ (right). The grid of emission positions is projected onto the plane of the accretion disk. The color bar is expressed in units of the ISCO orbital time $t_{\mathrm{ISCO}}$. The lensing effect of the black hole is clear for an inclination of $85^{\circ}$ on the right panel, leading to a higher concentration of emission points on the side of the black hole opposite to the observer. }
	\label{fig:hittime}
\end{figure*}

Before producing an image of the disk, one needs to compute the relativistic time delay for each point of the disk in order to take into account the multiple time steps of the fluid simulations that will correspond to the same observed time. Fig.~\ref{fig:hittime} shows the emission date of photons for two extreme inclinations (5$^{\circ}$ and 85$^{\circ}$). Fig.~\ref{fig:hittime} shows that the emission points are spread differently along the disk depending on the inclination. The distribution is homogeneous and isotropic for low inclinations, but is very inhomogeneous and anisotropic for high inclinations. This anisotropy is due to the projection effect due to the mapping of the observer's screen onto a very inclined disk, resulting in the emission points being aligned preferentially along the direction perpendicular to the line of sight. The inhomogeneity is due to the lensing effect of the black hole that concentrates emission points on the side of the black hole opposite to the observer. Fig.~\ref{fig:hittime} also clearly shows that the higher the inclination, the more time slices of data will be required. 
The respective difference between the maximal and minimal emission dates on the primary image are approximately $0.1\,t_{\mathrm{ISCO}}$ (left panel) and $0.9\,t_{\mathrm{ISCO}}$ (right panel). However, due to the strong beaming effect at high inclination, the bulk of the total specific intensity in one image comes from a small part of the disk. Thus, the emission dates of the photons reaching this small part of the disk are close to each other, and the final effect of time delay is only marginal on the light curve. We have checked that the difference between the exact light curve and a light curve computed without taking into account the time delay is at the level of one to a few percent only. 
\newline

Fig.~\ref{fig:image} shows the image (i.e. the map of specific intensity) of the accretion disk at an inclination\footnote{We call here inclination the angle between the $z$ axis and the line of sight. A null inclination thus corresponds to face-on view.} of 85$^{\circ}$, approximately at the time when the fundamental mode of the RWI dominates. 
\begin{figure}[htbp]
	\centering
	\includegraphics[width=7cm,height=7cm]{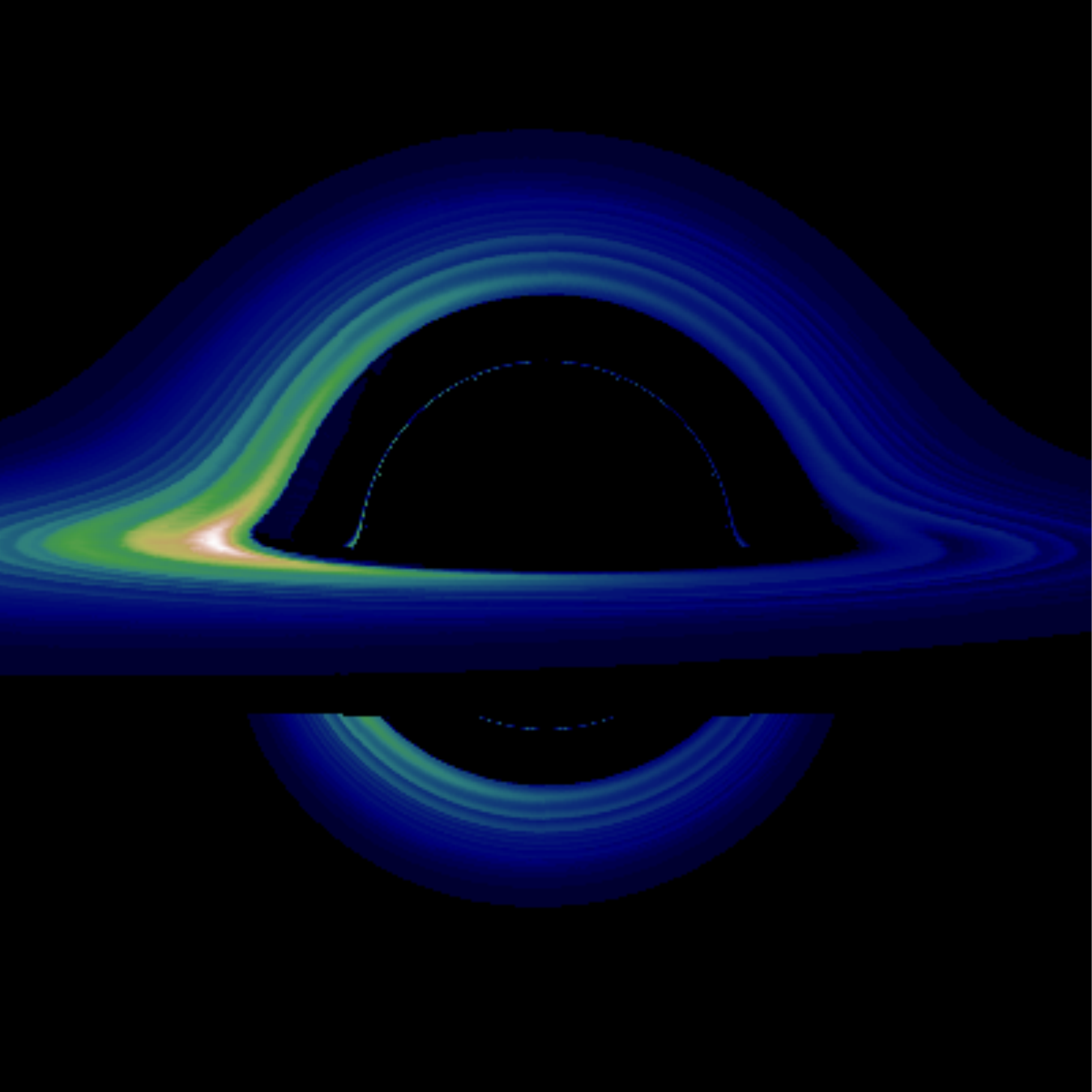}
	\caption{Image of a 2D optically thick disk emitting blackbody radiation with an inclination of 85$^{\circ}$ and in the presence of the RWI . The first order image (the distorted complete ring) clearly shows the spiral shape of the emitting region. The beaming effect makes the emission brighter on the approaching side of the disk (here, on the left of the image). The second order image is the portion of ring at the bottom of the image. It is due to photon swirling around the black hole before reaching the observer. The third order image is the thin ring of illuminated pixels at the center of the image, due to photons orbiting around the black hole very close to the event horizon before reaching the observer. This ring is the so-called black hole silhouette, it is truncated here due to optical thickness.}
	\label{fig:image}
\end{figure}
Each pixel of the image is obtained by interpolating between the set of simulated data at the time of emission of the photon, as stressed in Sect.~\ref{sec:raytracing}. Here, around $40$ different data time slices are used for computing the image. The instability has started to grow and one can identify the spiral density waves of the RWI. 
\begin{figure}[htbp]
	\centering
	\includegraphics[width=7cm,height=7cm]{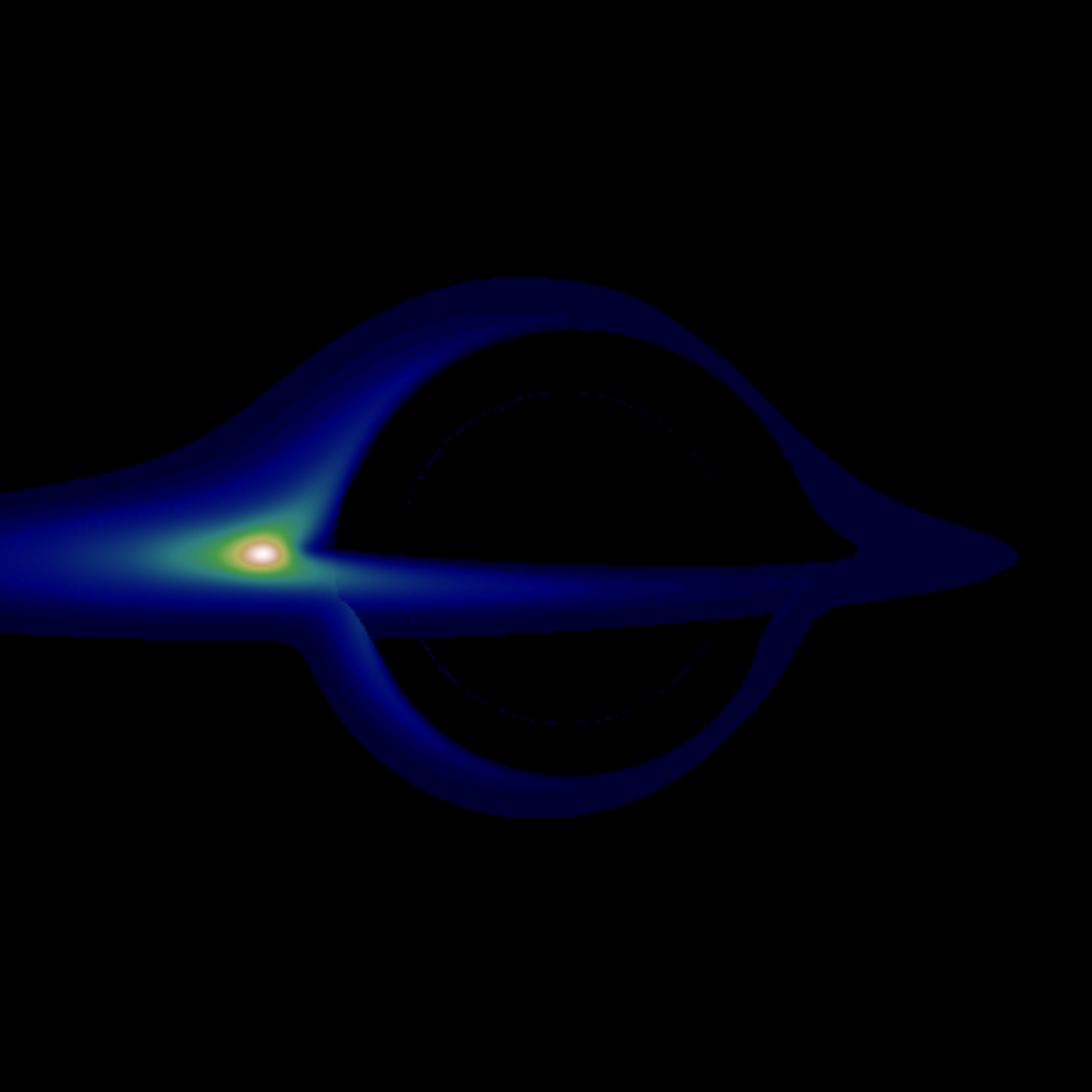}
	\caption{Image of a 3D disk emitting Bremsstrahlung radiation with an inclination of 85$^{\circ}$ and in the presence of the RWI. The overall emission is much more concentrated in the vicinity of the Rossby vortex as compared to Fig.~\ref{fig:image}.}
	\label{fig:image3D}
\end{figure}
On top of this modulation due to the instability, there is a clear beaming effect: matter moving towards the observer is brighter (here, on the left side of the image). The secondary image of the disk is visible as a semicircle at the centre of the primary image.

Fig.~\ref{fig:image3D} depicts the image of a 3D disk subject to the RWI, approximately at the time when the fundamental mode of the RWI dominates. Here, around $6$ different data time slices are used for computing the image: the time sampling has been reduced as 3D data are much heavier than 2D data. We checked at 2D that this reduced sampling still allows us to retrieve similar results.
As compared to Fig.~\ref{fig:image}, the 3D disk's emission is much more concentrated on the inner parts of the disk. This is due to the much stronger dependency of the 3D emission on temperature as compared
to the 2D case, as explained in Sect.~\ref{compare2D3D}. As the hottest parts of the disk are concentrated close to the radius of launching of the instability (see Fig.~\ref{fig:midplane}), only these inner regions are seen.


\section{Modulation of the light curve by the RWI}
\label{sec:results}

This section presents and analyzes the light curves obtained by ray-tracing the RWI hydrodynamical simulations. 
These are computed by summing the specific intensities over all solid angles on the observer's sky. 
This boils down to summing the disk images over all pixels, for all observation times.


\subsection{2D case: high time resolution}
\label{sec:2D}

    We first analyze the main characteristics of the light curve modulation in the 2D case where we can have a much better time resolution between the hydrodynamical snapshots\footnote{This is due to the smaller size of the 2D data files compared to the 3D.}.
  In order to catch the details of the light curve evolution, 40 frames of fluid simulations are computed during each period of rotation at ISCO. The ray-tracing code then interpolates between these fluid simulation results in order to compute the specific intensity map at different times of observation.  We compared the results obtained with the one from a higher  (80 frames) and smaller (20 frames and 6 frames) time resolution.
  While the overall behaviour was similar, 40 frames allow us to get a more detailed light curve, similar to that with 80 frames. We therefore decided to do all the runs in 2D at 40 frames per period of rotation at ISCO. 
   We computed the light curve from the moment the RWI started to grow in the disk until it reaches its non-linear states as shown by the amplitude evolution in Fig. \ref{fig:growth}.
\newline

At zero inclination, the light curve will slowly drift towards smaller values due to the black hole's accretion during the simulation. In order to determine the flux fluctuation that is only due to the RWI, it is important to subtract this continuous drift from the light curve. This is done by simply subtracting the light curve obtained at 1$^{\circ}$ of inclination from every other light curve. As the GYOTO code uses Boyer-Lindquist spherical-like coordinates, the $z$-axis is singular, and it is not possible to derive a light curve at exactly zero inclination. However, the light curve is only marginally impacted by beaming at 1$^{\circ}$ as compared to exactly 0$^{\circ}$. The error introduced by subtracting the light curve at 1$^{\circ}$ of inclination is a small underestimation of the amplitude of the modulation that does not impact our result here. Indeed, we are only looking for a proof of principle that the RWI could modulate the flux at a level compatible with the observed HFQPO. 

\begin{figure*}[htbp]
	\centering
	\includegraphics[width=7cm,height=7cm]{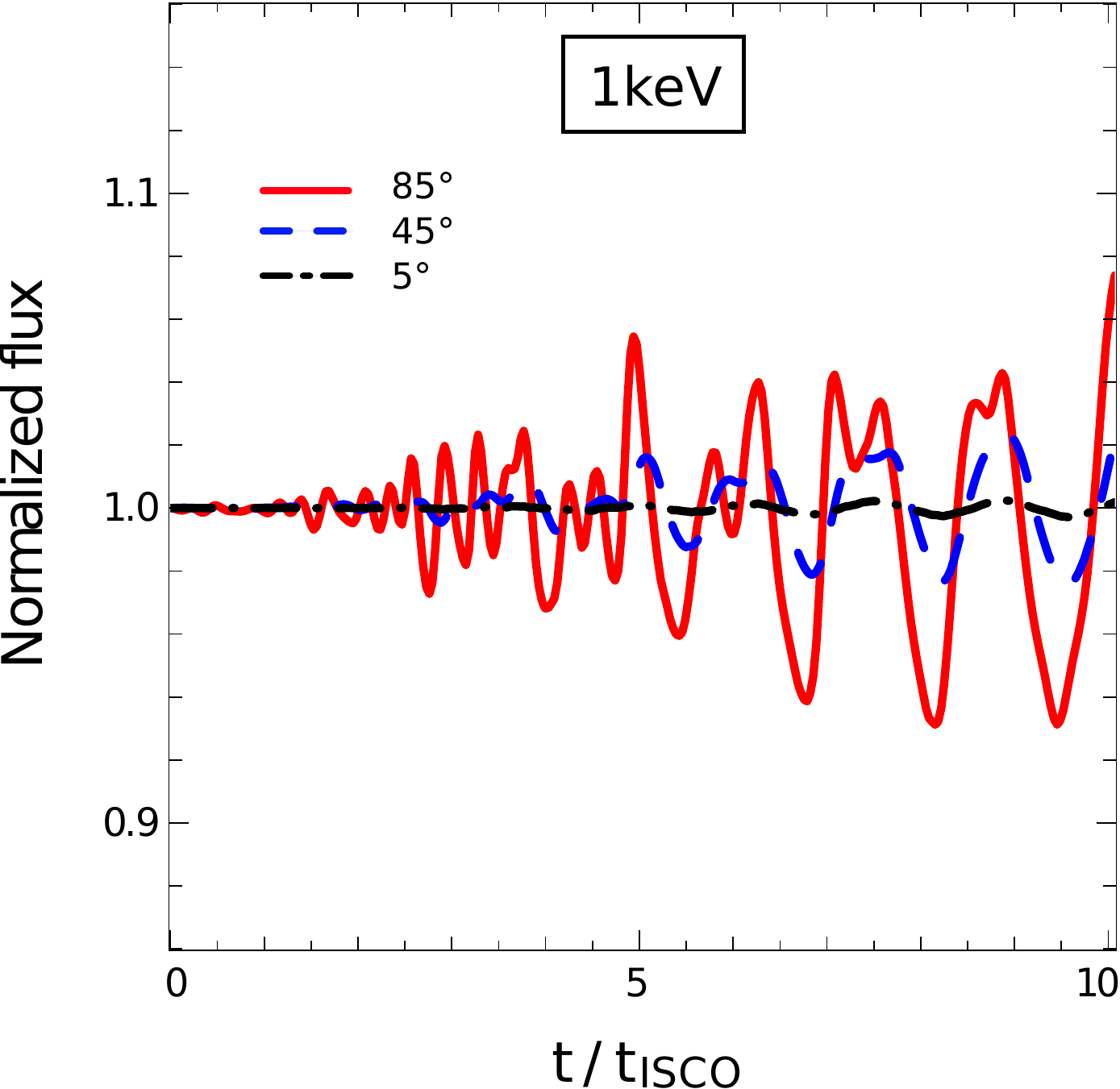}
	\hspace{1cm}
	\includegraphics[width=7cm,height=7cm]{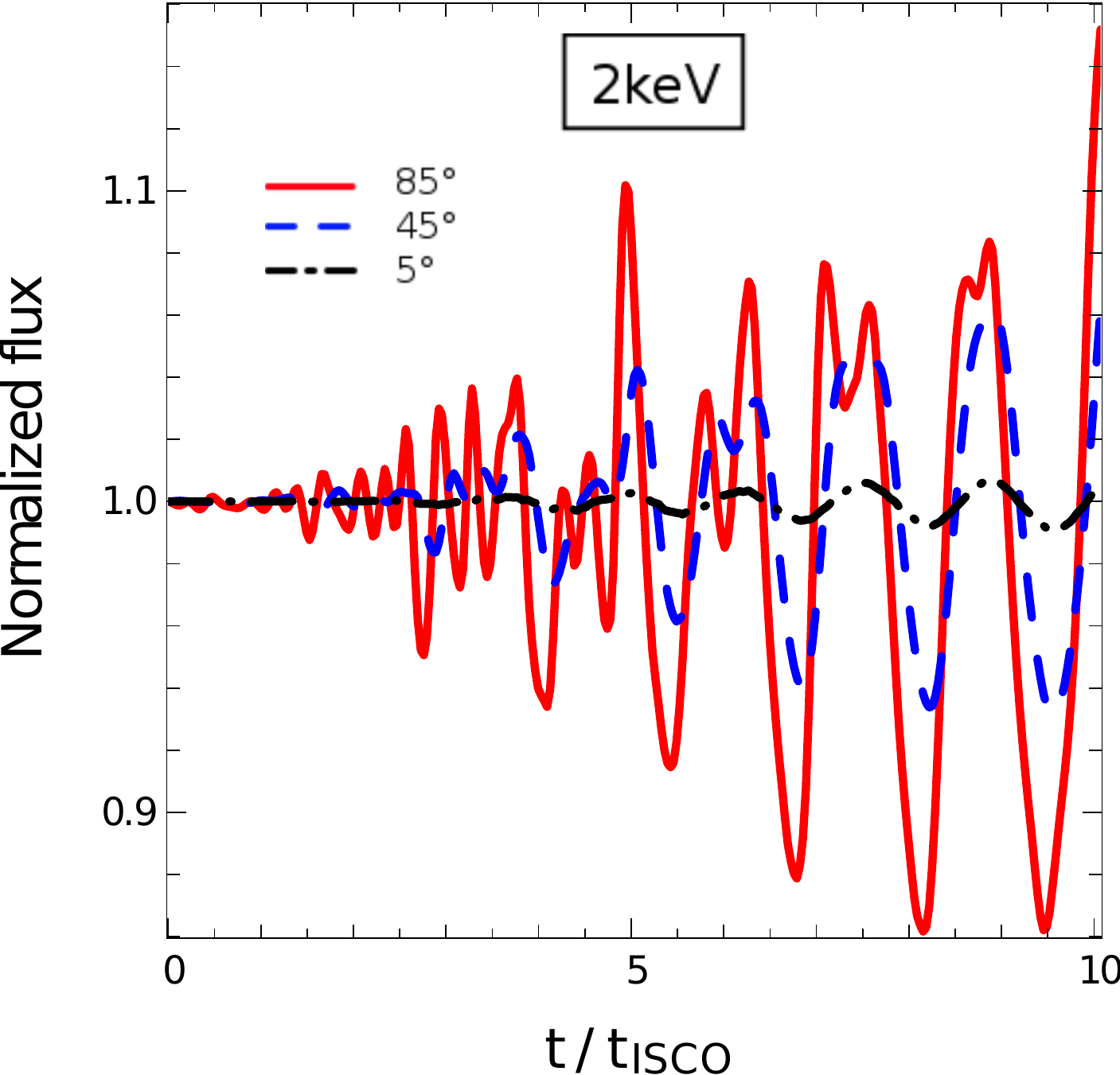}
	\caption{Light curves of a microquasar subject to the 2D RWI at 1~keV (left) and 2~keV (right) at different inclinations. The flux is normalized to its initial value. The inclination is 5$^{\circ}$ (dash-dotted black), 45$^{\circ}$ (dashed blue), or 85$^{\circ}$ (solid red).}
	\label{fig:2DLC}
\end{figure*}

Fig.~\ref{fig:2DLC} shows the light curves obtained for different values of the inclination parameter at two different values of the energy of the observed photon: $1$ and $2$~keV. Initially when the instability has a very low amplitude, the modulation of the flux is very low. After a few ISCO rotations, the instability is growing exponentially and modulates the flux at a detectable level. The period $T_{\mathrm{mod}}$ of the modulation equals the period of rotation of the Rossby vortices, i.e. the period of rotation at $r\approx1.4\,r_0$ (see Fig.~\ref{fig:criterium}), that is to say $T_{\mathrm{mod}}\approx1.6\,t_{\mathrm{ISCO}}$. 

Due to the general relativistic beaming effect, when the Rossby vortex is on the approaching side of the disk, the resulting flux is boosted at high inclination, the opposite being true for the receding side of the disk. 
This beaming effect is also visible in the light curve substructures that can be observed at high inclination. 
For instance, an $m=2$ mode will lead to two sharp peaks in the light curve at high inclination, related to the passage of the Rossby vortices at the approaching side of the disk. If the inclination is low, this effect is much fainter, leading to smaller substructures in the light curve. The oscillation frequency of the light curve evolves during the simulation with a high frequency after a few rotations. After $5$ rotations, a mixture of modes between one frequency and twice this frequency, can be identified. Whereas at the end of the simulation the mode with the lowest frequency dominates. This evolution corresponds to the evolution of modes seen in Fig.~\ref{fig:growth}: the dominant mode has initially a frequency $3\omega$, then $2\omega$ and eventually the fundamental mode dominates.

\begin{figure*}
	\centering
	\includegraphics[width=7cm,height=7cm]{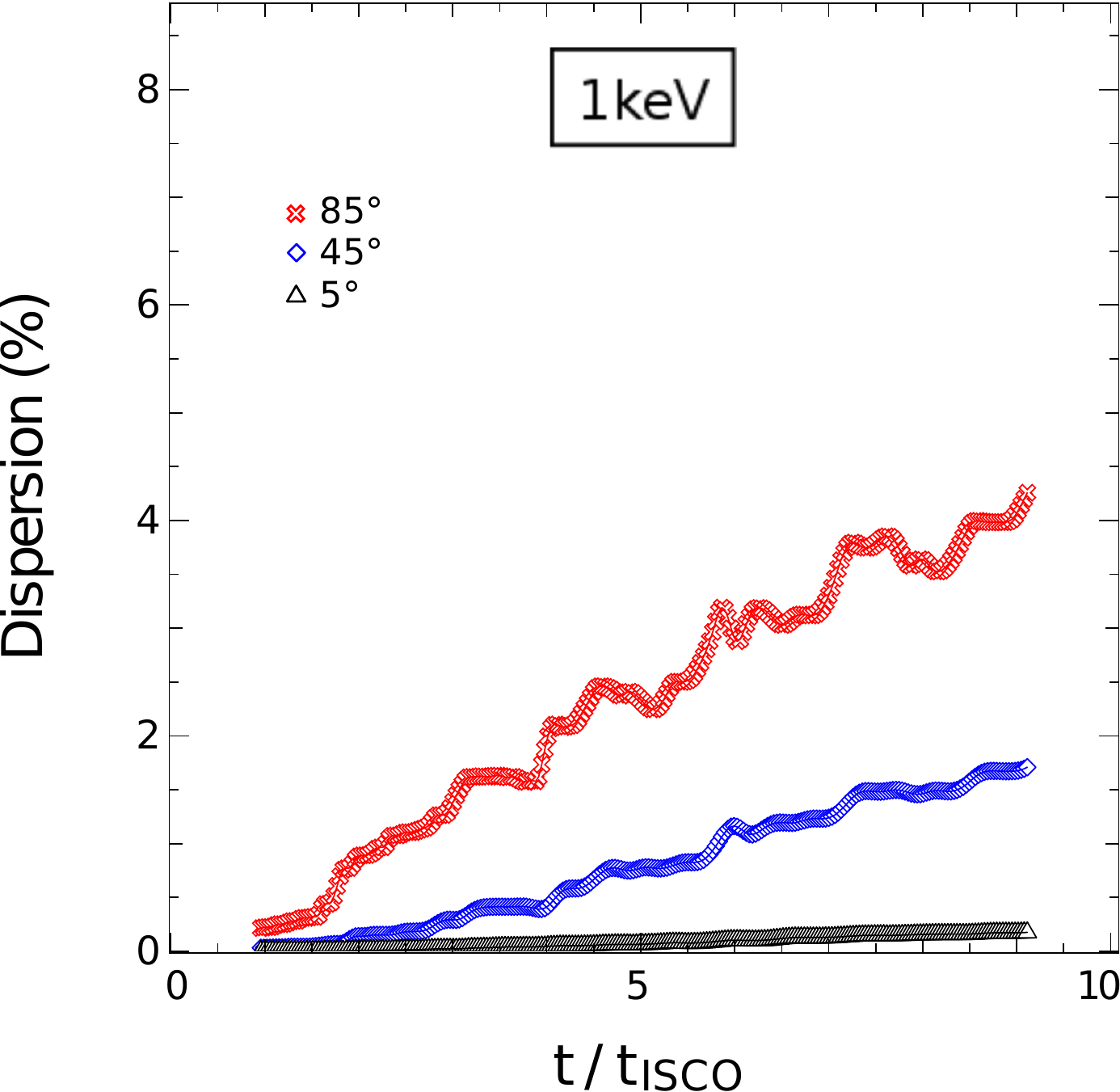}
	\hspace{1cm}
	\includegraphics[width=7cm,height=7cm]{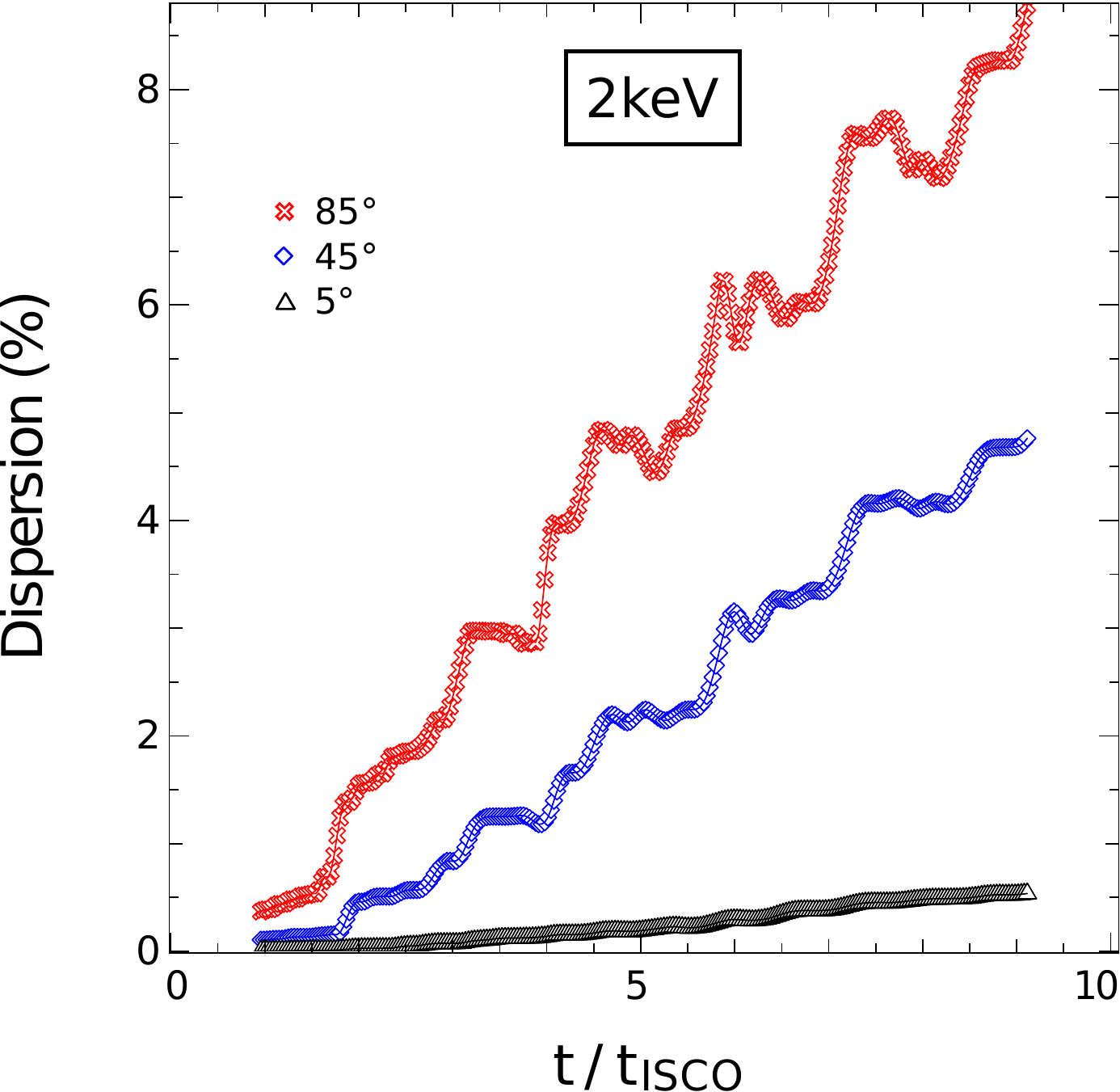}
	\caption{Dispersion of the light curve points in Fig.~\ref{fig:2DLC} at 1~keV (left) and 2~keV (right). Each point is obtained by computing the dispersion of the light curve points over two orbital periods. The inclination is 5$^{\circ}$ (black triangles), 45$^{\circ}$ (blue diamonds), or 85$^{\circ}$ (red crosses).}
	\label{fig:2DRms}
\end{figure*}

The comparison between the two values of energy of the observed photon shows that similar shapes are obtained for the two cases, but a higher modulation amplitude is reached for higher energy as expected due to the Planck law. Indeed, the ISCO temperature being $10^{7}$~K, the maximum frequency of the Planck law is closer to 2~keV that to 1~keV. 

Fig.~\ref{fig:2DRms} shows the light curve rms computed with a sliding window with a width of two orbital periods. 
The amplitude of the modulation is growing linearly with a change of slope at around $5\,t_{\mathrm{ISCO}}$, corresponding to the saturation of the instability (see Fig.~\ref{fig:growth}).
 The maximum level of modulation is of around $4\%$ at 1~keV and $8\%$ at 2~keV. This is somewhat higher than most observations of HFQPOs.
 \refe{ However, the maximum's exact value is influenced by the initial {$\cal{L}$} profile (see Eq.~\ref{eq:vortensity} and Fig.~\ref{fig:criterium}). Here we are only interested in seeing if a reasonable setup can modulate the flux to a level similar to the one observed. We keep the detailed study of the growth rate and saturation level for a forthcoming paper (Meheut, Lovelace, Lai, 2013) } 

\subsection{Full 3D case}
\label{sec:3D}

   We then turned to 3D simulation to see if there was any difference in the observables. There, we could not get a similar time resolution as in 2D
   because of the hydrodynamical data size. We therefore used the 40 frames per orbit resolution for only one period for confirmation while we made
   a longer term lightcurve at the much lower time resolution of 6 frames per orbit at the ISCO. As we will see below, this lower resolution is 
   still able to catch the broad aspects of the light curve, although it will not allow us to obtain the finest details.
    
     Fig.~\ref{fig:LC3D} depicts the light curve and its rms for a 3D disk subject to the RWI, for photons of observed energy equal to 2~keV, and for three different values of inclination. 
     The first important result is that the RWI is capable of modulating the light curve, similar to what had already been shown in the 2D case, 
     with a modulation of a few percent. This is a strong argument that makes the RWI a reliable model of microquasars HFQPOs.
	
	The domination of the different modes is also easily seen in Fig.~\ref{fig:LC3D}, with the mode $m=2$ dominating at the beginning
	of the simulation, and eventually the mode $m=1$. Here, the mode $m=1$ dominates after around $15$ ISCO periods, whereas it dominates after around
	$30$ periods in Fig.~\ref{fig:growth3D}. This difference is due to the fact that the ray-tracing simulations are initiated when the instability is already strong enough
	to give a non negligible rms. As a consequence, the $15$ first ISCO periods of the 3D hydrodynamics computations 
	where not used as their rms is extremely low. 
	
	The right panel of Fig.~\ref{fig:LC3D} shows the same characteristics as in the equivalent 2D Fig.~\ref{fig:2DRms}. The rms shows 
	a linear profile with a clear change of slope around $t\approx7\,t_{\mathrm{ISCO}}$. This is linked to mode $m=1$ starting to dominate over mode $m=2$.

This being given, there are two main differences between the 2D and 3D results:
\begin{description} 
\item[1-] The 3D modulation grows more slowly than its 2D counterpart.  
\newline
     This is comes from the differences in initial disk conditions, i.e. mainly to the choice of the density profile: different density profiles give \refe{different shapes for the extremum 
     of ${\cal L}$ which in turns give} 
     different growths of the instability. Figs.~\ref{fig:growth} and~\ref{fig:growth3D} already showed that the 3D growth rate is 
     slower than its 2D counterpart. This translates directly to the light curve (moreover, as stated above, the 3D light curve 
     initial point is $15$ ISCO periods after the launch of the instability).
     
     Let us stress that the \refe{index of the} density profile 
     is not  constrained by observations. As the difference of growth time depends on the choice of \refe{the power-law index}, 
     it cannot not be taken as an intrinsic difference between 2D and 3D RWI. 
                         

\item[2-] The light curve exhibits a slight asymmetry in the 3D case (the oscillation is not exactly centered on the initial flux value).
\newline
	This could be explained by the strong dependency on temperature of the 3D emission, as explained in Sect.~\ref{compare2D3D}.
	During the simulation, the temperature of the Rossby vortex increases by a few \%, due to accretion. Its emission thus becomes greater and greater. 
	This translates to a slight drift of the light curve maxima towards higher values. 
\end{description}

    

\begin{figure*}
	\centering
	\includegraphics[width=7cm,height=7cm]{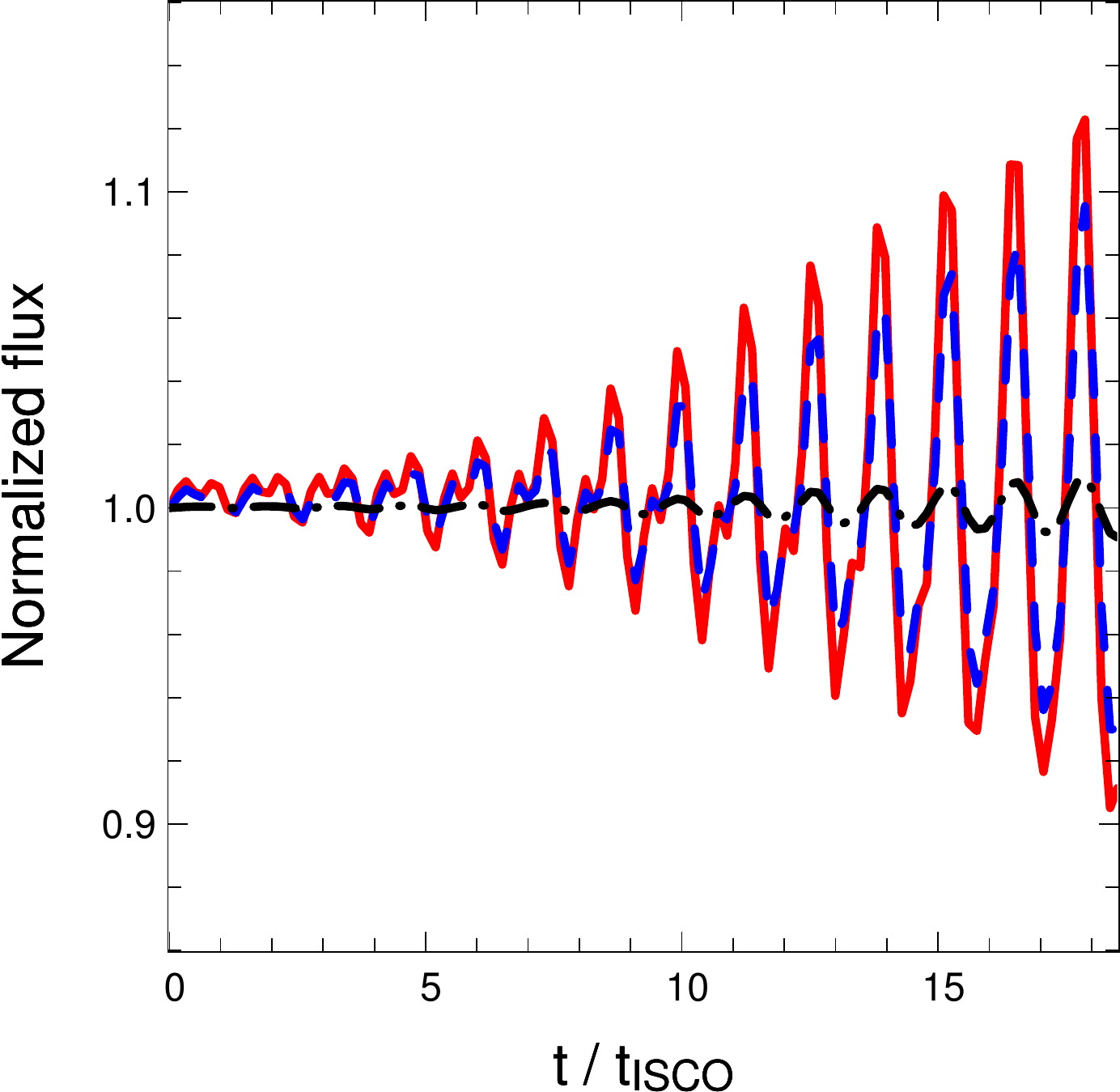}
	\hspace{1cm}
	\includegraphics[width=7cm,height=7cm]{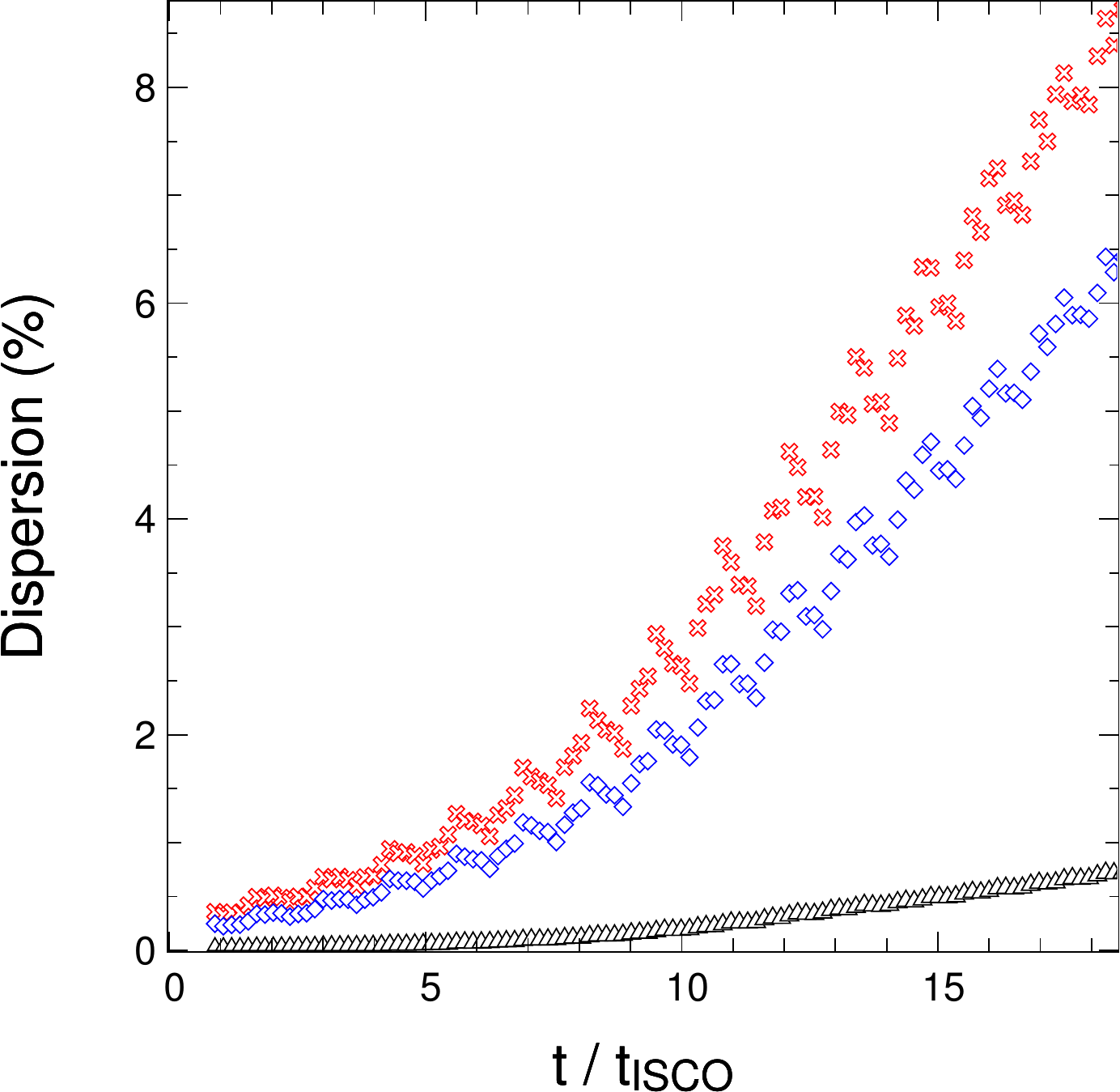}
	\caption{\textit{Left}: Light curves of a microquasar subject to the 3D RWI at 2~keV at an inclination of 5$^{\circ}$ (dash-dotted black), 45$^{\circ}$ (dashed blue) or 85$^{\circ}$ (solid red). \textit{Right}: Dispersion of the light curve points of the left panel. Each point is obtained by computing the dispersion of the light curve points over two orbital periods. The inclination is 5$^{\circ}$ (black triangles), 45$^{\circ}$ (blue diamonds) or 85$^{\circ}$ (red crosses). For both panels, mind the different time scale as compared to Figs.~\ref{fig:2DLC} and~\ref{fig:2DRms}.}
	\label{fig:LC3D}
\end{figure*}

\subsection{Are 3D simulations necessary to compare with observations?}
\label{sec:comp23}

Comparing Figs.~\ref{fig:2DLC} to~\ref{fig:LC3D}, it appears first that the light curve is modulated both in 2D and 3D, which is the main result of this paper. The dependency of the light curve as a function of the inclination parameter and frequency of the radiation is also very similar. As these general features of the 2D and 3D computations are alike, an interesting conclusion is that 2D results are sufficient in order to analyze the general observable characteristics of the RWI, at least until we get a better constraint on the profiles in the disk.

This is particularly interesting when considering the difference in terms of computing time and memory resources between a 2D and 3D simulation. The hydrodynamical data at a given time is around $100$ times heavier at 3D than 2D (typically respectively $50$~Mo and $0.5$~Mo).The time needed to compute an image from 3D data is typically $15$ times what is needed for a 2D computation.

However, let us stress that the present work does not allow us to compare the detailed time evolution of the 3D RWI  with the 2D case, 
due to our choice of time sampling in the ray-tracing computations (this choice being dictated by the computing time and memory resources 
needed for the 3D simulations). 
As the 3D RWI displays specific features in the $z$ direction~\citep{MEH10}, it may be that specific observable characteristics could be 
obtained only by resorting to high time resolution 3D simulations. 
Nevertheless, these features would be small corrections to the general trend that stays close to the 2D results, 
and would not be within reach of current instruments. Indeed the 2D time sampling of 40 frames per ISCO period
implies a time resolution of around $10^{-4}$~s for the observed light curve. This is far beyond current instrumental capabilities.

Moreover, let us stress that the radiative processes and radiative transfer are treated in a very simplified way in this study for the 3D case. The 2D simulations are thus sufficient 
only when one is not interested in studying precisely the radiative properties of the disk.




\section{Conclusion}
\label{sec:conclu}
 
  The RWI has been previously proposed as a model for HFQPOs and we have now demonstrated its ability to modulate the flux coming from
  the disk. Using 2D and 3D hydrodynamical simulations we have also been able to study how the amplitude of this modulation evolves 
  as a function of the source inclination and of the radiation frequency. The 2D simulations have been shown to be sufficient in order to recover
  the broad characteristics of the light curve.
  
  By using hydrodynamical simulations we were able to focus only on the RWI  and its effects on the light curve. If this can be considered for the
  case where HFQPOs occur alone in a softer state, as is the case for the $67$Hz modulation of GRS $1915$+$105$ for example~\citep{morgan97},
  HFQPOs are more commonly observed in the steep power law or hard intermediate state simultaneously with a LFQPO.
   In~\citet{VAR12a} we have demonstrated, in 2D, the ability of the RWI to co-exist with another instability that could give rise to the LFQPO.
   Our future work will be devoted to that particular case and in particular how it influences the observables. 
   Indeed, this state is much more frequent during microquasar outbursts than the softer state we studied here.
   


\begin{acknowledgements}
This work has been financially supported by the French GdR PCHE and Campus Spatial Paris Diderot. Some of the simulations were performed using HPC resources from GENCI-CINES (Grant 2012046810).
\end{acknowledgements}

\bibliography{biblio}

\end{document}